\documentclass[lettersize,journal]{IEEEtran}
\usepackage{amsmath,amsfonts}
\usepackage{algorithmic}
\usepackage{array}
\usepackage[caption=false,font=normalsize,labelfont=sf,textfont=sf]{subfig}
\usepackage{textcomp}
\usepackage{stfloats}
\usepackage{url}
\usepackage{verbatim}
\usepackage{graphicx}
\usepackage{float}
\restylefloat{figure}
\usepackage{pgfplots}
\pgfplotsset{compat=1.18} 
\usepackage{amsmath,xcolor}
\usepackage{authblk}

\expandafter\def\expandafter\normalsize\expandafter{%
    \normalsize%
    \setlength\abovedisplayskip{0pt}%
    \setlength\belowdisplayskip{8pt}%
    \setlength\abovedisplayshortskip{-8pt}%
    \setlength\belowdisplayshortskip{2pt}%
}

\def\BibTeX{{\rm B\kern-.05em{\sc i\kern-.025em b}\kern-.08em
    T\kern-.1667em\lower.7ex\hbox{E}\kern-.125emX}}
\usepackage{balance}
\begin{document}
\title{Real-Time Adaptive Neural Network on FPGA: Enhancing Adaptability through Dynamic Classifier Selection}
\author{Achraf El Bouazzaoui, Abdelkader Hadjoudja, Omar Mouhib
\thanks{Achraf El Bouazzaoui, Abdelkader Hadjoudja, Omar Mouhib are with the Laboratory of Electronic Systems, Information Processing, Mechanics and Energy, Ibn tofail University, BP 242, Kénitra, Morocco, email : achraf.elbouazzaoui@uit.ac.ma .}}

\maketitle

\begin{abstract}
This research studies an adaptive neural network with a Dynamic Classifier Selection framework on Field-Programmable Gate Arrays (FPGAs). The evaluations are conducted across three different datasets. By adjusting parameters, the architecture surpasses all models in the ensemble set in accuracy and shows an improvement of up to 8\% compared to a singular neural network implementation. The research also emphasizes considerable resource savings of up to 109.28\%, achieved via partial reconfiguration rather than a traditional fixed approach. Such improved efficiency suggests that the architecture is ideal for settings limited by computational capacity, like in edge computing scenarios. The collected data highlights the architecture's two main benefits: high performance and real-world application, signifying a notable input to FPGA-based ensemble learning methods.

\end{abstract}


\section{Introduction}
Neural networks\cite{bramer2020introduction}, a branch of machine learning\cite{choi2020introduction}, lead current computational research. They excel by discerning intricate non-linear patterns in data. Mirroring the human brain's neural structures, these systems are made up of interconnected nodes or neurons. These nodes cooperate in processing and conveying data. A system's complexity, assessed by its hidden layers and number of neurons, directly impacts its effectiveness and versatility for diverse roles. Such networks find applications in various sectors like business\cite{tkavc2016artificial}, renewable energy\cite{marugan2018survey}, healthcare\cite{shahid2019applications}, and more\cite{abiodun2018state}, indicating their essential role in recent problem solving.

The effectiveness of neural networks is offset by their considerable computational demands, which escalate as structural intricacy grows. This has accelerated the quest for ideal computational platforms. Conventional systems, such as CPU-based architectures and GPUs, have provided foundational support. but there is a shift toward reconfigurable architectures\cite{dhilleswararao2022efficient,vestias2019survey,wang2017reconfigurable}. Defined by their aptitude for post-fabrication alterations, these configurations deliver a synthesis of performance and adaptability. They facilitate hardware-level modifications, preserving the sort of flexibility usually linked with software environments.

Field-Programmable Gate Arrays (FPGAs) serve as prime representations of such reconfigurable systems. Their innate parallel processing capability, paired with immediate configuration options, designates them as a suitable neural network foundation . FPGAs promise tailored neural network frameworks\cite{dhilleswararao2022efficient}, optimizing both operational speed and energy conservation\cite{seng2021embedded}. 

In the advancing field of FPGA-based dynamic reconfigurable neural networks implementation, multiple pivotal research works have surfaced, each tackling specific issues and presenting novel solutions.
the research \cite{seyoum2020spatio} centered on the adaptation of Binary Neural Networks (BNNs) for FPGAs, particularly when faced with resource constraints. Examining their findings, a method grounded in mixed-integer linear programming became evident. This method, when melded with dynamic partial reconfiguration\cite{vipin2018fpga}, facilitated a streamlined BNN implementation on a Zynq-7020 FPGA.

In parallel, another research \cite{shi2023efficient} unveiled the dynamic reconfigurable CNN accelerator (EDRCA) structure, targeting the challenges posed by hardware limitations in embedded edge computing environments. By harnessing dynamic reconfiguration capabilities specific to FPGAs, the EDRCA adeptly navigated these challenges, introducing a methodology designed to enhance efficiency while minimizing resource utilization.

Finally, the work by \cite{khalil2022reconfigurable} presented a novel design approach tailored for economical neural network functionalities on FPGA. Central to this approach were "flexible" layers, crafted to fulfill dual functions, leading to a reduction in layer count. Evaluations conducted on an Altera Arria 10 FPGA indicated a noteworthy decrease in resource consumption, with these innovative layers maintaining precision and adaptability across varied applications.

In the current research, we investigate the role of dynamic classifier selection in enhancing classification accuracy within a hardware implementation context. Unlike previous studies, which primarily focus on partial reconfiguration tailored to specific tasks or applications, our approach leverages partial reconfiguration based on the unique characteristics of each data point. This method produces a system that is both highly adaptive and precisely tuned, effectively balancing accuracy with resource efficiency. Our findings advance the existing body of knowledge by presenting a novel data-driven strategy for improving system adaptability and optimization.

\section{Overview of DCS Module design}
\subsection{Dynamic Classifier selection : Definition}
\noindent 

Dynamic Classifier Selection (DCS) \cite{cruz2018dynamic,giacinto2000dynamic} serves as an essential mechanism specifically designed to enhance classification accuracy. Distinct from static ensemble methods, which combine outputs from various classifiers using pre-defined algorithms, DCS functions in an adaptive manner. The structure excels in managing issues like concept drift fluctuations \cite{gama2014survey,almeida2018adapting} in the statistical attributes of the target variable—and data imbalance \cite{johnson2019survey}, characterized by a skewed distribution of classes. This makes DCS particularly effective in fluctuating and intricate settings, providing a degree of adaptability and exactitude not commonly reached by static ensemble methods. 
In the context of this paper, the ensemble consists entirely of neural network variants. For each incoming test instance, DCS assesses the competence of each neural network model and selects the one most likely to provide an accurate classification. This process unfolds in four key stages: initial training of the neural network models, real-time competence evaluation for each test instance, dynamic selection of the most suitable neural network, and final classification of the test instance.

\begin{figure}[H]
    \includegraphics[width=0.47\textwidth]{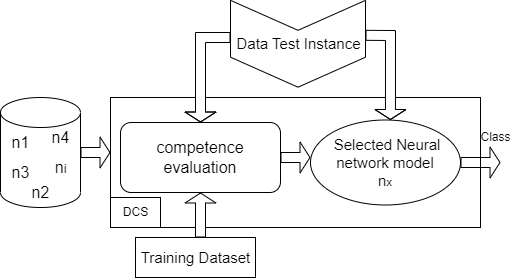}
    \caption{Diagram of dynamic classifier selection}
    \label{fig:nn_dcs.png}
\end{figure}

In this adapted version of Dynamic Classifier Selection (DCS), the method incorporates k-Nearest Centroid for estimating model competence\cite{soares2006using}. The feature space undergoes partitioning through the application of the k-means algorithm. Subsequent to this, every identified cluster is correspondingly linked to a neural network model, deemed most capable for the classification tasks pertinent to that specific cluster. Upon the introduction of a new test instance, an immediate assignment to its closest cluster occurs, followed by the invocation of the associated neural network model for the task of classification. This strategy, which centralizes around k-NC, optimally balances computational economy with classification precision.
\raggedbottom
\subsection{design of Competence Evaluation module}
\noindent 

The architecture of our dynamic classifier selection is predicated on an efficient k-Nearest Centroid (k-NC) implementation. Figure 2 elucidates the essential elements of this architecture: Centroid Read-Only Memory (ROM), Distance Modules (DM), Control Unit and Comparator.

\begin{figure}[H]
    \includegraphics[width=0.50\textwidth]{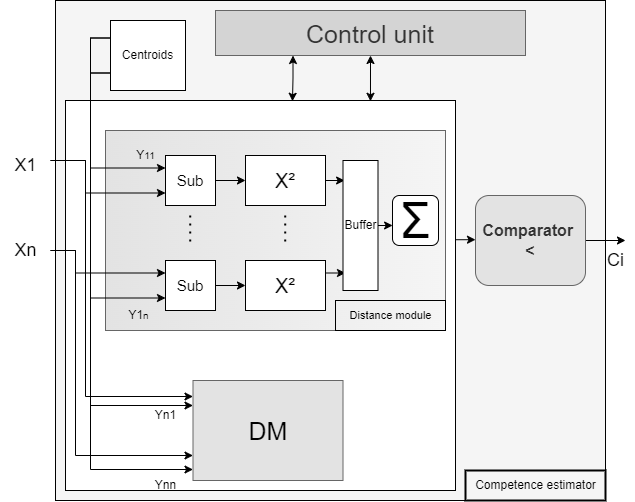}
    \caption{Diagram of the competence estimator}
    \label{fig:knc.png}
\end{figure}

Within each DM, subtractors work in parallel, along with a multiplier and an accumulator. These components are for distance computations between an input data vector and pre-stored centroids.

The subtractors within a single DM function in parallel, generating the deltas between individual features of the input vector and those of a given centroid. Subsequently, a multiplier squares these differences. An accumulator sums the products, thereby calculating the Squared Euclidean Distance pertaining to that centroid. The Euclidean distance is essentially calculated using the equation.

\begin{equation}
d(x, C) = \sqrt{\sum_{i=1}^{n} (x_i - C_i)^2}
\end{equation}
where $\mathbf{xi}$ and $\mathbf{Ci}$ are the individual features of the input vector and centroid.

The Control Unit is a critical component, managing both data flow and computational tasks. With only two Distance Modules available, the Control Unit orchestrates multiple cycles to evaluate all centroids relative to the input vector. It is responsible for reallocating centroids to the Distance Modules as needed, while also ensuring proper timing and precise data transfer between the different modules.

Upon the completion of centroid evaluation, the Control Unit channels the computed distances to a Distances Buffer for temporary storage. At this juncture, a Comparator queries the Distances Buffer to ascertain the least distance from among the calculated metrics. The Comparator then outputs the label associated with this minimal distance, thus determining the optimal accelerator for the processed input vector.

The Control Unit calculates distances to all centroids and stores them in a Buffer. A Comparator then examines all the stored distances, identifying the smallest one. It outputs a label corresponding to this minimum distance. This label indicates the most suitable associated accelerator for the input vector.

\section{ The neural networks models design}

In the devised neural network architecture,  the emphasis is on creating a parameterized architecture that allows for a variable number of neurons and hidden layers, This design is particularly tailored to accommodate hidden layers that utilize Multiply-Accumulate (MAC) computations and Rectified Linear Unit (ReLU) activations.

\begin{figure}[H]
    \includegraphics[width=0.45\textwidth]{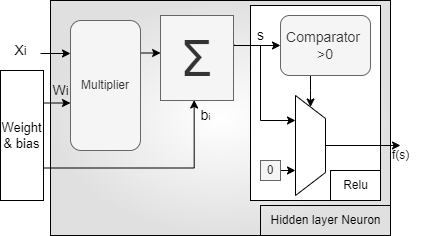}
    \caption{Diagram of the hidden layer neuron}
    \label{fig:neuron.png}
\end{figure}

Our FPGA-based neural network architecture is designed for flexibility, allowing parametrized numbers of neurons and hidden layers. Each neuron in the hidden layers performs a Multiply-Accumulate (MAC) operation, followed by a Rectified Linear Unit (ReLU) activation function.

Within the FPGA-based layout, each neuron in the hidden layers is precisely designed for executing a MAC operation on its array of input values.
the neuron multiplies each input element $\mathbf{X}$ with a corresponding weight $\mathbf{W}$ and subsequently adds a bias term $\mathbf{b}$, mathematically represented as\begin{equation}s = \sum_{i=1}^{n} w_i \cdot x_i + b\end{equation}

After the MAC operation, a Rectified Linear Unit (ReLU) activation function adds non-linearity to the neural model. In hardware, ReLU is implemented using a comparator and a 2:1 Multiplexer (MUX). The comparator evaluates the MAC output $\mathbf{s}$ and produces a binary control signal. This signal directs the MUX to output either the original MAC value or zero, effectively implementing the ReLU function: \begin{equation} y = \max(0, s)\end{equation}

In the final output layer, we use an argmax function instead of the typical softmax activation. This choice prioritizes classification accuracy over detailed class probability distributions. The argmax identifies the neuron with the highest output and converts it to a one-hot encoded vector. This approach reduces computational complexity, simplifies hardware implementation, and speeds up inference

\begin{figure}[H]
    \includegraphics[width=0.48\textwidth]{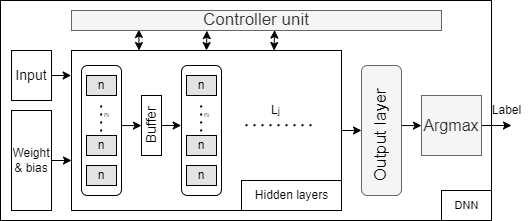}
    \caption{Diagram of the Neural network component}
    \label{fig:dnn.png}
\end{figure}

Figure 4 illustrates the structure of various neural network models in our ensemble. Each constituent model differs based on two fundamental parameters: neuron count and layer count. This  variance serves dual purposes. Primarily, it augments ensemble diversity, an essential element for the robustness of the Dynamic Classifier Selection mechanism deployed on FPGAs. Additionally, this diversity is strategically designed to amplify system accuracy, constituting a substantial advantage over reliance on a singular neural network configuration.
Equipped with a diverse ensemble of neural network models, the adaptative neural network possesses the capability to dynamically allocate the most suitable model for individual test instances using partial reconfiguration.

\begin{figure}[H]
    \centering
    \includegraphics[width=0.3\textwidth]{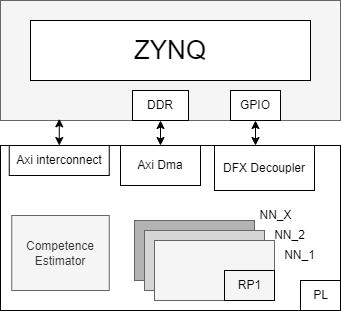}
    \caption{Diagram of the real-time Adaptive Neural Network Design}
    \label{fig:nn_arch.png}
\end{figure}

The foundation of our architecture is constituted by the Processing System (PS) incorporated into the Zynq System on Chip (SoC). As delineated in Figure 5,the PS is crucial in managing the operations of individual neural network modules as well as the competence estimator. This management is made possible due to the inherent high-performance, adaptability, and energy-efficient characteristics of Zynq platforms.
Data transfer between our neural networks and competence estimator occurs via the AXI Direct Memory Access (DMA) and AXI Interconnect systems. The AXI DMA plays a vital role in facilitating efficient data movement between system memory and neural network implementations. Furthermore, the AXI Interconnect operates using the AXI4-Lite protocol, which is instrumental in enabling the architecture to interpret output data result from both the accelerators and the competence estimator.
One feature of our architectural design is the incorporation of a Reconfigurable Partition (RP) region. Neural networks implementations are situated within this RP area, interfacing with the FPGA through a Dynamic Function eXchange (DFX) decoupler. Control over the DFX decoupler is exerted by an GPIO.

\section{Experimental Implementation and Analysis }
\subsection{Dataset and parameters}
In the current research, the Dynamic Classifier Selection (DCS) system's performance and adaptability, deployed on Field-Programmable Gate Arrays (FPGAs), undergo evaluation using three datasets. Distinctive in their levels of complexity, these datasets also diverge in their categorization tasks: one is oriented toward multiclass challenges and the other two focus on binary classification. Consistency in methodology is maintained by adopting an 85-15 data partitioning scheme. Specifically, 85\% of each dataset contributes to model training, with the remaining 15\% set aside for evaluation.
During data preparation, we split the training data into two distinct sets. The first set trained three neural network models, while the second was used the remaining two. Such a division aimed to increase the diversity in the ensemble set.

\begin{table}[H]
    \begin{tabular}{lcccc}
        \hline
        NN model & Lj(Vehicle) & Lj(Diabetes) & Lj(German Credit) \\
        \hline
        NN1 & (18,18,10)  & (5,3) & (7,7) \\
        NN2 & (30,30,20) & (3,3) &(7,7,4) \\
        NN3 & (27,27,22) & (12,12,8) & (4,4) \\
        NN4 &(20,20,15) & (8,8,4) &  (8,8,4) \\
        NN5 &(20,20,16) & (6,4,4) &  (6,6,4) \\
        \hline
        \hspace{1cm}
    \end{tabular}
    \caption{Neural Network Model Parameters (Layers and Neurons) Used}
    \label{tab:nn_vehicle}
\end{table}
\raggedbottom
Table I presents the customized neural network models for each of the three respective datasets, focusing on distinctions in the number of neurons and architectural layers.
In the performance evaluation module of the architecture, the ideal number of centroids depends on the specific dataset. For the Vehicle Silhouette and German Credit datasets, 70 centroids provide the best results. In contrast, the Diabetes dataset performs optimally with 50 centroids. These dataset-specific centroid counts are crucial for fine-tuning the competence estimator module.
\raggedbottom

\subsection{Ressource utilization}

Resource utilization stands as a pivotal metric in evaluating the efficiency of the dynamic architecture designed for FPGAs. In this analysis, we closely examine the resources consumed by each of the five neural network models for the vehicle data and the competence estimator.

\begin{table}[H]
\centering
    \begin{tabular}{lcccc}
        \hline
        NN models & LUTs & FFs & DSPs & BRAMs \\
        \hline
        NN1  & 4159  & 4658 & 100 &0\\
        NN2 & 6666 & 7137 &168 &0\\
        NN3 & 6428 & 6849 & 160 &0\\
        NN4  &4820 & 5309 &  118 &0\\
        NN5  &4849 & 5381 &  120 &0\\
        C.E.  &2958 & 6335 &  50 &7\\
        \hline
        \hspace{1cm}
    \end{tabular}
    \caption{Comparison of resources used by various neural networks}
    \label{tab:nn_vehicle2}
\end{table}
\raggedbottom

For a more exhaustive evaluation, a hypothetical scenario is also considered, wherein a static architecture is employed, and all neural network models are implemented concurrently.

\begin{table}[H]
\centering
    \begin{tabular}{lcccc}
        \hline
        Resource & Static Implementation & Dynamic Implementation \\
        \hline
        LUTs & 45397 (64.34\%) & 14224 (20.16\%)\\
        FFs & 49006 (34.73\%) & 17265 (12.23\%)\\
        DSPs & 716 (198.89\%) & 210 (58.33\%)\\
        BRAMs & 13  (6.02\%) & 8 (3.70\%)\\
        \hline
        \hspace{1cm}
    \end{tabular}
    \caption{Resource utilization: Comparison between Static and Dynamic Architectures}
    \label{tab:Resource utilization}
\end{table}
\raggedbottom

This resource evaluation is done on the Ultra96-V2 Board, which constitutes the hardware basis of our dynamic architecture. By contrasting resource utilization in both dynamic and static configurations, the objective is to quantify the efficiency improvements attributable to the dynamic approach. Such an assessment is of particular importance in the context of FPGA deployments, where resource optimization is integral to achieving both high performance and cost efficiency. Through this analysis, the aim is to corroborate the resource efficiency of the dynamic architecture, Resource utilization was notably minimized, up to a factor of 109.28\%, thereby affirming its applicability that demand computational efficiency.

\section{Results}

Following an in-depth exploration of parameter tuning, specifically concerning the number of centroids for the competence estimator and the configurations of hidden layers in the neural network models, we now pivot to the results. This subsequent analysis leverages the optimized parameters to provide a quantitative assessment of the real-time adaptive neural network.

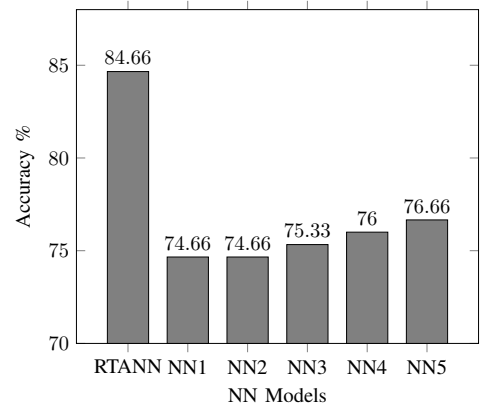
\begin{figure}[H]
\centering
\caption{Performance evaluation for german credit dataset}
\begin{tikzpicture}[scale=0.78]
    \begin{axis}[
        ybar,
        bar width=20pt,
        enlarge x limits={abs=25pt},
        ymin=70, ymax=88,
        ylabel={Accuracy \%},
        xlabel={NN Models},
        symbolic x coords={RTANN,NN1, NN2, NN3, NN4, NN5},
        xtick=data,
        nodes near coords,
        nodes near coords align={vertical},
    ]

    \addplot[fill=gray]  coordinates {(RTANN, 84.66) (NN1, 74.66) (NN2, 74.66) (NN3, 75.33) (NN4, 76.00)(NN5, 76.66)};
    \end{axis}
\end{tikzpicture}
\end{figure}

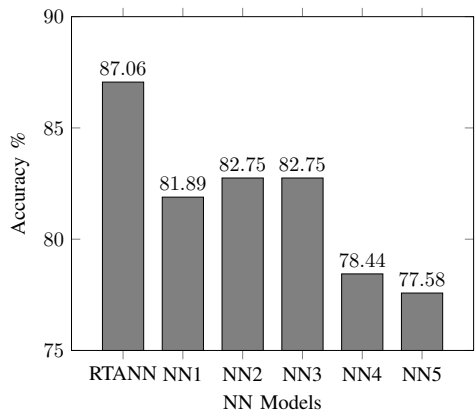
\begin{figure}[H]
\centering
\caption{Performance evaluation for diabetes dataset}
\begin{tikzpicture}[scale=0.78]
    \begin{axis}[
        ybar,
        bar width=20pt,
        enlarge x limits={abs=25pt},
        ymin=75, ymax=90,
        ylabel={Accuracy \%},
        xlabel={NN Models},
        symbolic x coords={RTANN,NN1, NN2, NN3, NN4, NN5},
        xtick=data,
        nodes near coords,
        nodes near coords align={vertical},
    ]

    \addplot[fill=gray]  coordinates {(RTANN, 87.06) (NN1, 81.89) (NN2, 82.75) (NN3, 82.75) (NN4, 78.44)(NN5, 77.58)};
    \end{axis}
\end{tikzpicture}
\end{figure}

\begin{figure}[H]
\centering
\caption{Performance evaluation for vehicle dataset}
\begin{tikzpicture}[scale=0.78]
    \begin{axis}[
        ybar,
        bar width=20pt,
        enlarge x limits={abs=25pt},
        ymin=82, ymax=97,
        ylabel={Accuracy \%},
        xlabel={NN Models},
        symbolic x coords={RTANN,NN1, NN2, NN3, NN4, NN5},
        xtick=data,
        nodes near coords,
        nodes near coords align={vertical},
    ]

    \addplot[fill=gray]  coordinates {(RTANN, 94.48) (NN1, 85.04) (NN2, 85.04) (NN3, 84.25) (NN4, 88.18)(NN5, 88.97)};
    \end{axis}
\end{tikzpicture}
\end{figure}
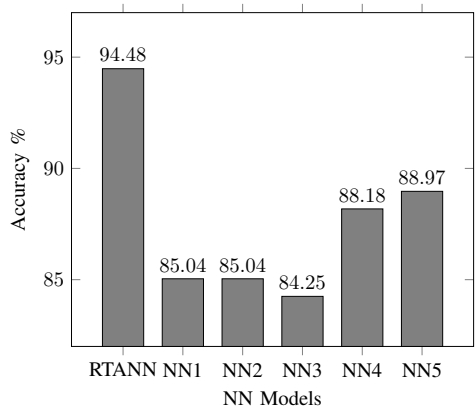

Figures 6, 7, and 8 illustrate a numerical comparison between a real-time adaptive neural network, based on a DCS architecture, and the individual neural network models that make up the ensemble. The data shows that, for the German Credit dataset, the DCS-based architecture achieved an accuracy rate of 84.66\%, which is 8\% higher than the most accurate individual model, which recorded 76.66\%. Similarly, for the Diabetes dataset, the DCS-based architecture attained an accuracy of 87.06\%, surpassing the leading individual model’s accuracy of 82.75\%. Lastly, in the Vehicle Silhouette dataset, the DCS-based architecture achieved an accuracy of 94.48\%, significantly outperforming the highest-scoring individual model, which had an accuracy of 88.97\%.

\section{Conclusion}

In conclusion, this study provides a comprehensive evaluation of a real-time, adaptive neural network using a DCS-based structural framework, tested across three diverse datasets. By optimizing parameters, the architecture consistently outperformed individual ensemble models in classification accuracy. Resource usage was significantly reduced by up to 109.28\% through partial reconfiguration compared to a static implementation. These findings confirm the high performance of the architecture and demonstrate its practical applicability in scenarios where balancing accuracy and resource efficiency is crucial.

\bibliographystyle{ieeetr}
\bibliography{nn.bib}

\end{document}